\documentclass{llncs}

\usepackage{cite}
\usepackage[pdftex]{graphicx}
\usepackage[usenames,dvipsnames]{xcolor}
\DeclareGraphicsExtensions{.pdf,.jpeg,.png}
\usepackage{amsmath,amsfonts,amssymb}
\usepackage{bbm}
\newcommand\numberthis{\addtocounter{equation}{1}\tag{\theequation}}
\usepackage{mathrsfs}
\usepackage{dsfont}
\usepackage{amsbsy}
\usepackage{stmaryrd}
\usepackage{url}
\usepackage{tikz}
\usetikzlibrary{arrows,positioning,automata, backgrounds}

\newcommand{\upsize}{^{(N)}}

\newcommand{\class}{\mathscr{A}} 
\newcommand{\cstsp}{S} 
\newcommand{\ctrsp}{E} 
\newcommand{\lbset}{\mathscr{L}} 

\newcommand{\loctr}[1]{\epsilon_{#1}} 
\newcommand{\loclab}[1]{\alpha_{#1}} 
\newcommand{\locarr}[1]{s_{#1} \xrightarrow{\loclab{i}} s'_{#1}} 

\newcommand{\sa}{Y} 

\newcommand{\pop}{\mathcal{X}} 
\newcommand{\pstsp}{\mathcal{S}} 
\newcommand{\ptrsp}{\mathcal{T}} 
\newcommand{\pstvec}{\mathbf{X}} 
\newcommand{\pinst}{\mathbf{x}_0} 
\newcommand{\syncset}{\mathbb{S}}
\newcommand{\upvec}{\mathbf{v}}

\newcommand{\drift}{\textbf{F}} 
\newcommand{\fluid}{\boldsymbol\Phi} 

\newcommand{\dta}{\mathbb{D}}
\newcommand{\atprop}{\Gamma_{\pstsp}} 
\newcommand{\gstsp}{Q}
\newcommand{\ginst}{q_0}
\newcommand{\gfinst}{F}
\newcommand{\gtr}{\rightarrow}

\newcommand{\lmulti}{\{\hspace{-0.8mm}\lvert}
\newcommand{\rmulti}{\lvert\hspace{-0.8mm}\}}

\newcommand{\martr}{\mathcal{M}}
\newcommand{\dettr}{\mathcal{E}}

\newcommand{\gen}{\mbox{\textbf{G}}}

\newcommand{\mypar}[1]{\vspace{1ex}\noindent\textbf{#1.}\ }

\begin{document}


\title{ Fluid Model Checking of Timed Properties\thanks{ This research has been partially funded by the EU-FET project QUANTICOL (nr. 600708) and by  the German Research Council (DFG) as part of the Cluster of Excellence on Multimodal Computing and Interaction at Saarland University.}}

\author{Luca Bortolussi\inst{1} \and Roberta Lanciani\inst{2}}

\institute{
Modelling and Simulation Group, Saarland University, Germany\\ 
DMG, University of Trieste, Italy\\
CNR/ISTI, Pisa, Italy\\
\email{luca@dmi.units.it}
\and
IMT Lucca, Italy\\
\email{roberta.lanciani@imtlucca.it}}

\titlerunning{Fluid Model Checking of Timed Properties}
\authorrunning{L. Bortolussi \& R. Lanciani}
\maketitle


\begin{abstract}
We address the problem of verifying timed properties of Markovian models of large populations of interacting agents, modelled as finite state automata. In particular, we focus on time-bounded properties of (random) individual agents specified by Deterministic Timed Automata (DTA) endowed with a single clock. Exploiting ideas from fluid approximation, we  estimate the satisfaction probability of the DTA properties by reducing it to the computation of the transient probability of a subclass of Time-Inhomogeneous Markov Renewal Processes with exponentially and deterministically-timed transitions, and a small state space. For this subclass of models, we show how to derive a set of Delay Differential Equations (DDE), whose numerical solution provides a fast and accurate estimate of the satisfaction probability. In the paper, we also prove the asymptotic convergence of the approach, and exemplify the method on a simple epidemic spreading model. Finally, we also show how to construct a system of DDEs to efficiently approximate the average number of agents that satisfy the DTA specification. 

\vspace{3mm}
\end{abstract}

\pagenumbering{arabic}
\pagestyle{plain}



\section{Introduction}
\label{sec:intro}

One of the major technological challenges in computer science and engineering  is the design and analysis of large-scale distributed systems, where many autonomous components interact in an open environment. Examples include the public and shared transportation in smart cities, the power distribution in smart grids, and the robust communication protocols of online multimedia services. In this context, the mathematical and computational modelling plays a crucial role in the management of such \textit{Collective Adaptive Systems} (CAS), due to the need of understanding and control of their emergent behaviours in open working conditions. The intrinsic uncertainty of CAS can be properly captured by \textit{stochastic models}, but the large number of interacting entities always results in a severe \textit{state space explosion}, introducing exceptional computational challenges. In particular, the scalability of the models and of their analysis techniques is a major issue in the development of \textit{stochastic model checking} procedures for the verification of formal properties. In this context, up to now, the numerical approaches  \cite{prism} are deeply hampered by the state space explosion of the large stochastic models, and the statistical methods based on simulation require a large computational effort.

A recent line of work tries to address the issue of scalability by exploiting stochastic approximation techniques \cite{qest, epew}, like the \textit{Fluid Approximation} \cite{kurtz, FMC, tutorial}. In this method, a stochastic discrete model is replaced by a simpler continuous one, whose dynamics is described by a set of differential equations. In \cite{FMC}, the authors exploit this limit construction to verify properties that asses the behaviour of a single individual in a collective system, and define a procedure called the \textit{Fluid Model Checking} (FMC) \cite{qapl15,LLM13}. This technique is based on the \textit{Fast Simulation Theorem} \cite{darling}, which ensures that in a large population, a single entity is influenced only by the mean behaviour of the rest of the agents. 

In this work, we extend \cite{FMC} to more complex \textit{time-bounded properties} specified by \textit{Deterministic Timed Automata} endowed with a single clock \cite{alur,mcbook,cslta}. As in \cite{FMC,HaydenTSE,qest, katoen}, we combine the agent and the DTA specification with a product construction, obtaining a \textit{Time-Inhomogeneous Markov Renewal Process} \cite{cinlar}. We then exploit results \cite{qest12,hayden}, defining the Fluid Approximation of this type of models as the solution of a system of \textit{Delay  Differential Equations} (DDE) \cite{darling}. Other works dealing with the verification of DTA properties are \cite{fu, chen, chen2013, chen2011}.

\vspace{1mm}
\noindent\textbf{Main Result.}  We introduce a new fast and efficient Fluid Model Checking procedure to accurately approximate the probability that a single agent satisfies a single-clock DTA specification up to time $T$. Similarly to \cite{FMC}, the technique is based of the Fast Simulation Theorem, and couples the Fluid Approximation of the collective system with a set of Delay Differential Equations for the transient probability of the Time-Inhomogeneous Markov Renewal Process obtained by  the product construction between the single agent and the DTA specification.
\vspace{1mm}

In the paper, we discuss the \textit{theoretical aspects} of our approach, proving the \textit{convergence} of the estimated probability to the true one in the limit of an infinite population. We also show the procedure at work on a running example of a simple epidemic process, emphasising the quality of the approximation and the gain in terms of computational time. Finally, by exploiting the construction of \cite{qest,hayden}, we also show how to define a set of DDEs approximating the mean number of agents satisfying a single-clock DTA specification up to time $T$.

\vspace{1mm}
\noindent\textit{Paper structure.} In Sec. \ref{sec:background}, we introduce the modelling language, the Fluid Approximation, the Fast Simulation Theorem, and the DTA specification for the timed properties. In Sec. \ref{sec:FMC}, we present our FMC procedure, defining the DDEs for the probability that the single agent satisfies the timed property. In Sec. \ref{sec:Fluid}, we adapt our verification technique to compute the mean number of agents that meet the DTA requirement. In Sec. \ref{sec:ExpRes}, we discuss the quality of the approximation on the epidemic example. Finally, in Sec. \ref{sec:conc}, we draw the final conclusions. The proofs of the theoretical results are reported in the Appendix.
\vspace{3mm}


\section{Background and Modelling Language}
\label{sec:background}

\subsubsection{Agent Classes and Population Models.}
\label{subsec:PopMod}
A collective system is comprised of a large number of interacting \textit{agents}. To describe its dynamics, we define a \textit{population model} \cite{qest, epew} in which the agents are divided into classes, called \textit{agent classes}, according to their behaviour.

\begin{definition}[Agent Class]
\label{agentclass}
An agent class $\class$ is a pair $(\cstsp, \ctrsp)$ in which $\cstsp = \{1, \ldots, m\}$ is the (finite)  state space and $\ctrsp = \{\loctr{1}, \ldots, \loctr{\eta}\} \subseteq \cstsp\times \lbset \times \cstsp$ is the (finite) set of local transitions of the form $\loctr{i} = \locarr{i}$, where $s_i, s'_i \in \cstsp$ are the initial and arrival states, and $\alpha_i \in \lbset$ is the unique label of $\epsilon_i$, i.e. $\alpha_i\neq \alpha_j$ for $i\neq j$\footnote{The restriction on the uniqueness of the labels can be dropped (as in \cite{qest}) at the price of heavier notation and combinatorics in the definitions of the rest of the paper.}.
\end{definition}

Intuitively, an agent in class $\class = (\cstsp, \ctrsp)$ is a finite state automaton that can change state by performing the actions in $\ctrsp$. Then, assuming that agents in the same state are indistinguishable, to define the population model, we rely on the \emph{counting abstraction}, counting how many agents are in each state at time $t$. Hence, for each agent class, we define the \textit{collective} or  \textit{counting variables} $X\upsize_1(t), \ldots, X\upsize_m(t)$ given by $X\upsize_{j}(t) = \sum_{k} \mathds{1}_{\{Y\upsize_{k}(t) = j\}}$, where $Y\upsize_{k}(t) \in \{1, \ldots, m\}$ is the random variable denoting the state of agent $k$ at time $t$, and $N = \sum_\class \sum_j X\upsize_{j}$ is the \textit{population size}, that we assume constant in time (cf. also \cite{FMC}).  Then, given $n= \sum_\class \lvert\cstsp\lvert$, the state of the population model is given by the vector $\pstvec\upsize(t) \in (\mathds{R}_{\geq 0})^n$ that enlists the counting variables of the agent classes. 

\begin{definition}[Population Model] 
\label{populationModel}
A \textit{population model} $\pop\upsize$ is a tuple $\pop\upsize = (\mathbb{A}, \ptrsp\upsize, \pinst\upsize)$, where $\mathbb{A} = \{\class_1, \ldots, \class_\nu\}$ is the set of agent classes, as in Definition \ref{agentclass}; $\pinst\upsize = \pstvec\upsize(0)$ is the initial state; and $\ptrsp\upsize = \{ \tau_1, \ldots, \tau_\ell \}$ is the set of global transitions of the form $\tau_i = (\syncset_i, f\upsize_i, \upvec\upsize_i)$, where:
\begin{itemize}
\item $\syncset_i= \lmulti s_1 \xrightarrow{\loclab{1}} s'_1, \ldots, s_p \xrightarrow{\loclab{p}} s'_p \rmulti$ is the (finite) multi-set of local transitions synchronized by $\tau_i$;
\item $f\upsize_{i}: (\mathds{R}_{\geq 0})^n \longrightarrow \mathds{R}_{\geq 0}$ is the  (Lipschitz continuous) global rate function;
\item $\upvec_{i}=\sum_{\alpha_j \in \syncset_i}\lvert\lmulti \alpha_j\rmulti\lvert (\mathbbm{1}_{s_j} - \mathbbm{1}_{s'_j})$ is the update vector, where $\lvert\lmulti \alpha_j\rmulti\lvert$ is the multiplicity of $\alpha_i$ in $\syncset_i$, and $\mathbbm{1}_{s_j}$ is the vector equal to $1$ on $s_j$ and 0 elsewhere.
\end{itemize}
\end{definition}
When a global transition $\tau_i = (\syncset_i, f\upsize_i, \upvec_i)$ is taken, the transitions in $\syncset_i$ fire at the local level, meaning that, for each $s\xrightarrow{\alpha} s'$ in $\syncset_i$, an agent moves from $s$ to $s'$. Hence, the update vector $\upvec_i$ encodes the net change in the state $\pstvec\upsize(t)$ of $\pop\upsize$ due to transition $\tau_i$. Moreover, for the model to be meaningful, whenever at time $t$ it is not possible to execute $\tau_i$, because there are not enough agents available, i.e. $\left(\pstvec\upsize (t) - \upvec_i \right)_j < 0$ for some $j \in \{1, \ldots, n\}$ with $n = \lvert \pstvec\upsize (t) \lvert$, we require the rate function to be zero, i.e. $f\upsize_i(\pstvec\upsize (t)) = 0$.

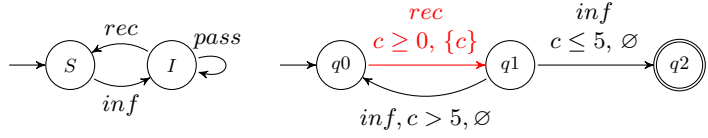
\begin{figure}[t!]
\begin{center}
\begin{small}
\hspace{-8mm}\begin{tikzpicture}[on grid, shorten <=1pt, >=stealth', auto, , scale=0.9]
	\node at (-1.4,0)   [state, draw=none] (q)   {};	
	\node at (1.5,0)   [state, scale=0.8] (I)   {$I$};
	\node at (0,0)   [state, scale=0.8] (S)   {$S$};
	\node at (2.6,0)   [state,draw=none] (Q)   {};
	\node at (4,0)   [state, scale=0.8] (q0)   {$q0$};
	\node at (6.5,0)   [state, scale=0.8] (q1)   {$q1$};
	\node at (9,0)   [state, accepting, scale=0.8] (q2)   {$q2$};
	
	\path [->]   (S)		edge [below, bend right]	node {$inf$}(I)
  						(I)		edge [above, bend right]	node {$rec$}(S)
  						(I)		edge [loop right, above, near start]	node {$pass$}(I)
  						(q) 		edge [] node{} (S)
  						(q0)		edge [above, color=red]	node {\begin{tabular}{c}$rec$\\$c\geq 0$, $\{c\}$\end{tabular}}(q1)
  						(q1)		edge [above]	node {\begin{tabular}{c}$inf$\\$c\leq 5$, $\varnothing$\end{tabular}}(q2)
  						(q1)		edge [below, bend left]	node {\begin{tabular}{c}$inf,c> 5,\varnothing$\end{tabular}}(q0)
  						(Q) 		edge [] node{} (q0);
\end{tikzpicture}
\end{small}
\end{center}
\caption{The agent class $\class$ (left) and property $\dta$ (right) of the running example.}
\label{fig:sis}
\end{figure}

\paragraph{Example.} The running example that we consider is a simple \textit{SIS model}, describing the spreading of a disease inside a population. All agents belong to the same agent class $\class$, depicted in Fig. \ref{fig:sis}, and can be either \textit{susceptible} ($S$) or \textit{infected} ($I$). When they are \textit{susceptible}, they can be infected (\textit{inf}), and when they are \textit{infected}, they can either pass the infection (\textit{pass}) or recover (\textit{rec}). Hence, the state $\pstvec\upsize(t)$ of the population model is $\pstvec\upsize(t) = (X\upsize_{S}(t), X\upsize_{I}(t))$, and we define 2 global transitions: $\tau_{r} = (\{I \xrightarrow{rec} S\}, f\upsize_{r})$ and $\tau_{i} = (\{S \xrightarrow{inf} I, I \xrightarrow{pass} I\}, f\upsize_{i})$. The former, $\tau_{r}$, mimics the recovery of one entity inside the population, while $\tau_{i}$ synchronises two local actions, namely $S \xrightarrow{inf} I$ and $I \xrightarrow{pass} I$, and models the transmission of the virus from an infected agent to a susceptible one. Finally, the rate functions depend on the number of agents involved in the transitions and follow the classical \textit{rule of mass action} \cite{andersson2000stochastic}: $f\upsize_{r}(t) = k_{r} X\upsize_{I}(t)$ and $f\upsize_{i}(t) = \frac{1}{N}k_{i}X\upsize_{S}(t)X\upsize_{I}(t)$, where $k_{r}, k_{i} \in \mathbb{R}_{\geq 0}$.

\subsubsection{Fluid Approximation.}
\label{subsec:FluidApprox}
The \textit{Fluid Approximation} \cite{kurtz, FMC, tutorial} of a population model $\pop\upsize= (\mathbb{A}, \ptrsp\upsize, \pinst\upsize)$ is an estimate of the \textit{mean} behaviour of its agents. To compute this approximation, we first \textit{normalise} $\pop\upsize$ by dividing the state vector $\pstvec\upsize(t)$ and the initial state $\pinst\upsize$ by the population size $N$, i.e. we define $\widehat\pstvec\upsize(t)= {\pstvec\upsize(t)}/{N}$ and $\widehat{\boldsymbol x}_0\upsize = {\pinst\upsize}/{N}$.
Then, for all transition $\tau_i \in \ptrsp\upsize$, we let $\widehat{f}_i\upsize(\widehat\pstvec)$ be the rate function, where we substitute the counting variables of $\pstvec\upsize(t)$ with the new normalised counting variables of $\widehat\pstvec(t)$. Moreover, we assume that for each $\widehat{f}_i\upsize(\widehat\pstvec)$, there exist a \textit{Lipschitz function} $f_i(\widehat\pstvec)$ such that $\widehat{f}_i\upsize(\widehat\pstvec)/N \xrightarrow{N\rightarrow +\infty} f_i(\widehat\pstvec)$ uniformly. Finally, in terms of $f_i(\widehat\pstvec)$, we define the \textit{drift} $\drift(\widehat\pstvec)$ given by $\drift(\widehat\pstvec)= \sum_{\tau_i}\upvec_if_i(\widehat\pstvec)$, whose components represent the instantaneous net flux of agents in each state of the model. Then, given a \textit{time horizon} $T<+\infty$, the \textit{Fluid Approximation} $\fluid(t)$ of $\pop\upsize$ is the unique\footnote{Existence and uniqueness of $\fluid(t)$ are guaranteed by the Lipschitzianity of the $f_i(\widehat\pstvec)$.} solution of the system of \textit{Ordinary Differential Equations} (ODEs) given by
$$
\frac{d\fluid}{dt}(t) = \drift(\fluid(t)),\qquad \mbox{for\ }0 \leq t \leq T,
$$
with $\fluid(0) = \pinst$. The accuracy of the approximation \textit{improves} the larger is the ensemble of agents that we consider, i.e. \textit{the larger is $N$}, and is exact in the limit of an infinite population. Indeed, the following theorem holds true\cite{kurtz}.
\begin{theorem}[Fluid Approximation] 
For any $T < + \infty$ and $\epsilon >0$,
$$
Prob\left\lbrace\sup_{0\leq t\leq T}\lvert\lvert\widehat\pstvec\upsize(t)- \fluid(t)\lvert\lvert > \epsilon \right\rbrace\xrightarrow{N\rightarrow  +\infty} 0.
$$
\end{theorem}
\subsubsection{Fast Simulation.}
\label{subsec:fastsim}
In this paper, we are interested in the behaviour of a (random) \textit{single agent} inside a population. As we have just seen,  the dynamics of a large population can be accurately described by a \textit{deterministic} limit, the Fluid Approximation. But when we focus on one single agent in a collective system, we need to keep in mind that its behaviour in time will always remain a \textit{stochastic} process, even in large populations. Nevertheless, the \textit{Fast Simulation Theorem} \cite{darling,boudec, gast} guarantees that in the limit of an infinite population size, the stochastic process of the single agent senses only the \textit{mean} behaviour of the rest of the agents (i.e. there is no need to keep track of all the states of all the other entities in the population). This means that, when the population size is large enough, to analyse the dynamics the single agent, we can define the Fluid Approximation of the population model, and then use its state (i.e. the mean state of the rest of the agents) to compute the rates of a \textit{Time-Inhomogeneous CTMC} (ICTMC) \cite{FMC} that describes the behaviour of the single agent. 

Formally, let $\sa\upsize(t)$ be the stochastic process that describes the state of the single agent in the population model $\pop\upsize = (\mathbb{A}, \ptrsp\upsize, \pinst\upsize)$ with state vector $\pstvec\upsize(t)$. By definition, $\sa\upsize(t)$ \textit{is not independent of} $\pstvec\upsize(t)$. Now consider the normalised model $\widehat{\pop}\upsize$ described by $\widehat\pstvec\upsize(t)$, and let $\fluid(t)$ be the Fluid Approximation of $\pop\upsize$. Define the \textit{generator matrix} $\boldsymbol Q\upsize(\boldsymbol x)  = (q\upsize_{ij}(\boldsymbol x))$ of $\sa\upsize(t)$ as a function of the normalised counting variables, i.e. $\forall\ q\upsize_{ij}(\boldsymbol x)$,
$$
Prob\left\lbrace\sa\upsize(t+dt)=j\ \lvert\ \sa\upsize(t) = i, \widehat\pstvec\upsize(t) = \boldsymbol x\right\rbrace = q\upsize_{ij}(\boldsymbol x)dt.
$$
Notice that $\boldsymbol Q\upsize(\boldsymbol x)$ can be computed from the rates in $\pop\upsize$. Indeed, for $i\neq j,$
$$
q\upsize_{ij}(\boldsymbol x) = \sum_{\tau\in \ptrsp}\left[\frac{\lvert\lmulti i \rightarrow j \in \syncset_\tau \rmulti\lvert}{X_i}\frac{\widehat{f}_\tau\upsize(\widehat\pstvec)}{N}\right],
$$
where $\lvert\lmulti i \rightarrow j \in\syncset_\tau \rmulti\lvert$ is the multiplicity of $i \rightarrow j$ in the transition set $\syncset_\tau$ of $\tau$, i.e. the number of agents that take such transition according to $\tau$. Furthermore, as customary, let $q\upsize_{ii}(\boldsymbol x)= - \sum_{j\neq i} q\upsize_{ij}(\boldsymbol x)$. Then, since $\widehat{f}_i\upsize(\widehat\pstvec)/N \xrightarrow{N\rightarrow +\infty} f_i(\widehat\pstvec)$, we have that $\boldsymbol Q\upsize(\boldsymbol x) \rightarrow \boldsymbol Q(\boldsymbol x)$, where $\boldsymbol Q(\boldsymbol x)$ is computed in terms of the Lipschitz limits $f_i(\widehat\pstvec)$. Now, define the stochastic processes:
\begin{enumerate}
\item $Z\upsize(t)$, that describes the state of the process $\sa\upsize(t)$ for the single agent in class $\class$, when $\sa\upsize(t)$ is marginalised from $\pstvec\upsize(t)$;
\item $Z(t)$, that is the ICTMC, defined on the same state space of $Z\upsize(t)$, with \textit{time-dependent} generator matrix $\boldsymbol Q(\fluid(t))$, i.e. the generator matrix $\boldsymbol Q(t)$, where the Lipschitz limits $f_i(t)$ are computed over the components of $\fluid(t)$.
\end{enumerate}
Then, the following theorem can be proved \cite{darling}.
\begin{theorem}[Fast Simulation]\label{FastSimTheorem}
For any time horizon $T < +\infty$ and $\epsilon >0$, 
$$
Prob\left\lbrace\sup_{0\leq t\leq T}\lvert\lvert Z\upsize(t)- Z(t)\lvert\lvert > \epsilon \right\rbrace\xrightarrow{N\rightarrow  +\infty} 0.
$$
\end{theorem}

\paragraph{Example.} For the running example, if we consider a population of 1000 agents, i.e $N=1000$, and an initial state $\pinst\upsize=(900,100)$, then the Fluid Approximation $\fluid(t)$ of the population model is the unique solution of the following ODEs:
\begin{equation}\label{eqFluidSIS}
\begin{cases}
\frac{d\Phi_S}{dt}(t) = - k_i\Phi_I(t)\Phi_S(t) + k_r\Phi_I(t);\\
\frac{d\Phi_I}{dt}(t) = + k_i\Phi_I(t)\Phi_S(t) - k_r\Phi_I(t);
\end{cases}
\quad\mbox{with}\quad
\begin{cases}
\Phi_S(0) = 0.9;\\
\Phi_I(0) = 0.1.
\end{cases}
\end{equation}
The generator $\boldsymbol Q(\fluid(t))$ of the ICTMC $Z(t)$ for the single agent, instead, is:
\begin{equation}\label{eqgenSIS}
q_{S,S} (t) = - q_{S,I}(t);\ \ \ 
q_{S,I} (t) = k_i\Phi_I(t);\ \ \ 
q_{I,S} (t) = k_r;\ \ \ 
q_{I,I} (t) = - q_{I,S}(t).
\end{equation}

\subsection{Timed Properties}
\label{subsec:properties}

We are interested in properties specifying how a single agent behaves in \textit{time}. In order to monitor such requirements, we assign to it a unique \textit{personal clock}, which starts at time $0$ and can be reset whenever the agent undergoes specific transitions. In this way, the properties that we consider can be specified by a \textit{single-clock Deterministic Timed Automata} (DTA)\cite{alur,katoen}, which keeps track of the behaviour of the single agent with respect to its personal clock. Moreover, since we want to exploit the Fast Simulation Theorem, we restrict ourselves to \textit{time bounded} properties and, hence, we assign to the DTA a finite \textit{time horizon} $T<+\infty$, within which the requirement must be true.  

\begin{definition}[Timed Properties]
\label{def:aDTA}
A timed property for a single agent in agent class $\class$ is specified as a single-clock DTA of the form $\dta = \dta(T) = (T, \lbset, c,  \mathcal{C}\mathcal{C},$ $\gstsp,\ginst, \gfinst, \gtr)$, where $T < +\infty$ is the finite time horizon; $\lbset$ is the label set of $\class$; 
$c$ is the personal clock; $\mathcal{C}\mathcal{C}$ is the set of clock constraints, which are conjunctions of atoms of the form $c<\lambda$, $c\leq \lambda$, $c\geq \lambda$ or $c>\lambda$ for $\lambda \in \mathbb{Q}$; 
$\gstsp$ is the (finite) set of states;
 $\ginst \in \gstsp$ is the initial state; 
 $\gfinst \subseteq \gstsp$ is the set of final (or accepting) states; 
 and $\gtr\ \subseteq \gstsp \times \lbset \times  \mathcal{C}\mathcal{C} \times \{\varnothing, \{c\}\}  \times  \gstsp$ is the edge relation. Moreover, $\dta$ has to satisfy:
\begin{itemize}
\item (determinism) for each initial state $q\in Q$, label $\alpha\in\lbset$, clock constraint $c_{\bowtie} \in \mathcal{C}\mathcal{C}$, and clock valuation $\eta(c)\in\mathbb{R}_{\geq 0}$, there exists exactly one edge $q \xrightarrow{\alpha, c_{\bowtie}, r}q'$ such that $\eta(c) \models_{\mathcal{C}\mathcal{C}} c_{\bowtie}$\footnote{The notation  $\eta(c) \models_{\mathcal{C}\mathcal{C}} c_{\bowtie}$ stands for the fact that the value of the valuation  $\eta(c)$ of $c$ satisfies the clock constraint $c_{\bowtie}$.};
\item (absorption) the final states are all absorbing.
\end{itemize}
\end{definition}

A timed property $\dta$ is assessed over the time-bounded paths (of total duration $T$) of the agent class $\class$ sampled from the stochastic processes $Z\upsize(t)$ and $Z(t)$ defined for the Fast Simulation in Sec. \ref{subsec:fastsim}. The labels of the transitions of $\class$ act as inputs for the DTA $\dta$, and the latter is defined in such a way that it \textit{accepts} a time-bounded path $\sigma$ if and only if the behaviour of the single agent encoded in $\sigma$ satisfies the property represented by $\dta$. Formally, a time-bounded path $\sigma = s_0 \xrightarrow{\alpha_0,t_0} s_1 \xrightarrow{\alpha_1,t_1} \ldots \xrightarrow{\alpha_n,t_n} s_{n+1}$ of $\class$ sampled from $Z\upsize(t)$ (resp. $Z(t)$), with $\sum_{j=0}^n t_j  \leq T$, is \textit{accepted} by $\dta$ if and only if there exists a path $q_0 \xrightarrow{\alpha_0} q^{(1)} \xrightarrow{\alpha_1} q^{(2)}\xrightarrow{\alpha_2}\ldots\xrightarrow{\alpha_n} q^{(n+1)}$ of $\dta$ such that  $q^{(n+1)}\in F$. In the path of $\dta$, $q^{(i+1)}\in Q$ denotes the (unique) state that can be reached form $q^{(i)}\in Q$ taking the action $q^{(i)} \xrightarrow{\alpha_{i}, c_{\bowtie}, r}q^{(i+1)}$ whose clock constraint $c_{\bowtie}$ is satisfied by the clock valuation $\eta(c)$ updated according to time $t_{i}$. In the following, we will denote by $\Sigma_{\class, \dta, T}$ the \textit{set of time-bounded paths of $\class$ accepted by $\dta$}.

\paragraph{Example.} We consider the following property for the running example: \textit{within time $T$, the agent gets infected at least once during the $\Delta=5$ time units that follow a recovery}. To verify such requirement, we use the DTA $\dta = \dta(T)$ represented in Fig. \ref{fig:sis}. If we record the actions of the single agent on $\dta$, i.e. we synchronise $\class$ and $\dta$, when the agent recovers ($rec$), $\dta$ passes from state $q0$ to $q1$, resetting the personal clock $c$. After that, if the agent gets infected ($inf$) within 5 time units, the property is satisfied, and $\dta$ passes from state $q1$ to $q2$, which is accepting. If instead the agent is infected ($inf$) after 5 units of time, $\dta$ moves back to state $q0$, and we start monitoring the behaviour of the agent again. In \textcolor{red}{red} we highlight the transition that resets the personal clock $c$ in $\dta$.



\section{Fluid Model Checking of Timed Properties}
\label{sec:FMC}

Consider a single agent of class $\class = (\cstsp,\ctrsp)$ in a population model  $\pop\upsize = (\class, \ptrsp\upsize, \pinst\upsize)$, and a timed property $\dta = \dta(T) = (T, \lbset, \atprop,  \mathcal{C}\mathcal{C}, \gstsp,\ginst, \gfinst, \gtr)$. Let $\Sigma_{\class, \dta, T}$ be the set of time-bounded paths of $\class$ accepted by $\dta$. Moreover, let $Z\upsize(t)$ and $Z(t)$ be the two stochastic processes defined for the Fast Simulation in Sec. \ref{subsec:fastsim}. The following result holds true.
\begin{proposition}
The set $\Sigma_{\class, \dta, T}$ is measurable for the probability measures\linebreak $Prob_{Z\upsize}$ and $Prob_{Z}$ defined over the paths of $Z\upsize(t)$ and $Z(t)$, respectively.$\ \ \qed$
\end{proposition}  
Let $P\upsize(T) = Prob_{Z\upsize} \{\Sigma_{\class, \dta, T}\}$ and $P(T) = Prob_{Z} \{\Sigma_{\class, \dta, T}\}$. In this paper, we are interested in the \textit{satisfaction probability} $P\upsize(T)$, i.e. the probability that the single agent satisfies property $\dta$ within time $T$ in $\pop\upsize$. Then, the main result that we exploit in our Fluid Model Checking procedure is that, when the population is large enough (i.e $N$ is large enough), $P\upsize(T)$ can be accurately approximated by $P(T)$, which is computed over the ICTMC $Z(t)$, whose rates are defined in terms of the Fluid Approximation $\fluid(t)$ of $\pop\upsize$. The correctness of the approximation relies on the Fast Simulation Theorem and is guaranteed by the following result.
\begin{theorem}
\label{th:convergence}
For any  $T < +\infty, \qquad \lim_{N\rightarrow\infty} P\upsize(T) = P(T).\qquad\qquad\qquad\qed$ 
\end{theorem}
Moreover, to compute $P(T)$, we consider a suitable product construction $\class_\dta = \class \otimes \dta$, whose state is described by a \textit{Time-Inhomogeneous Markov Renewal Process} (IMRP) \cite{cinlar} that we denote by $Z_{\class_\dta}(t)$. In the rest of this section, we define $\class_\dta$ and $Z_{\class_\dta}(t)$, and we show how to compute the \textit{satisfaction probability} $P(T)$ in terms of the \textit{transient probability} $\boldsymbol P (T)$ of $Z_{\class_\dta}(t)$.


\mypar{The Product $\class_\dta$} We now introduce the product $\class_\dta$ between $\class$ and $\dta$, whose state is described by  a \textit{Time-Inhomogeneous Markov Renewal Process} (IMRP) $Z_{\class_\dta}(t)$ that has rates computed over the Fluid Approximation $\fluid(t)$ of $\pop\upsize$.

A \textit{Markov Renewal Process} (MRP) \cite{cinlar} is a jump-process, where the sojourn times in the states can have a general probability distribution. In particular, in the MRP $Z_{\class_\dta}(t)$, we will allow both \textit{exponentially} and \textit{deterministically-timed} transitions, and in the following, we will refer to them as the \textit{Markovian} and \textit{deterministic transitions}, respectively. Since the transition rates of $Z_{\class_\dta}(t)$ will be time-dependent, $Z_{\class_\dta}(t)$ will be a \textit{Time-Inhomogeneous} MRP.

To define the product $\class_\dta =(\class, S_\dta, \{\mathcal{M}, \mathcal{E}\}, s_{0,\dta}, F_\dta)$, let  $\delta_1 < \ldots < \delta_k$ be the (ordered) constants that appear in the clock constraints of $\dta$, and extend the sequence with $\delta_0 = 0$ and $\delta_{k+1} = T$. 
The \textit{state space} $S_{\dta}$ of $\class_\dta$ is given by $\{1,\ldots, k+1\} \times S \times Q$. The first element of $S_{\dta}$ identifies a time region of the clock $c$, and we refer to $S_{\dta_i} = \{(i,s,q)\ \lvert\ s\in S, q\in Q\}$ as the $i$\textit{-th Time Region} of $S_\dta$. The rest of $\class_{\dta}$ will be defined in such a way that the agent is in $S_{\dta_i}$ if and only if $c$ satisfies $\delta_{i-1} \leq \eta(c) \leq \delta_i$, where $\eta$ is the valuation of $c$. 
 
The \textit{set $\martr$ of Markovian transitions} of $\class_{\dta}$ is the smallest relation such that
\begin{equation}
\label{martr}
\forall\ i \in {1,\ldots, k+1,}\quad \frac{s \xrightarrow{\alpha} s' \in E\ \wedge\ q \xrightarrow{\alpha, c_{\bowtie}, \varnothing}q' \in \rightarrow \ \wedge\ \ [\delta_{i-1}, {\delta_i}] \models c_{\bowtie} }{ (i, s, q) \xrightarrow{\alpha}(i, s', q') \in \martr},
\end{equation}
\begin{equation}
\label{restr}
\forall\ i \in {1,\ldots, k+1,}\quad \frac{s \xrightarrow{\alpha} s' \in E\ \wedge\ q \xrightarrow{\alpha, c_{\bowtie}, \{c\}}q' \in \rightarrow\ \wedge\ [\delta_{i-1}, {\delta_i}] \models c_{\bowtie}}{ (i, s, q) \xrightarrow{\alpha}(1, s', q') \in \martr}.
\end{equation}

Intuitively, rule (\ref{martr}) synchronises the local transitions $s \xrightarrow{\alpha} s' \in E$ of the agent class $\class = (S,E)$ with the transition $q \xrightarrow{\alpha, c_{\bowtie}, \varnothing}q' \in \rightarrow$ that has the same label in $\dta$, obtaining a local transition $(i, s, q) \xrightarrow{\alpha}(i, s', q') \in \martr$ in $\class_\dta$ for each time region $i$ whose time interval $[\delta_{i-1}, \delta_i] \subseteq [0,T]$ satisfies the clock constraint $c_{\bowtie}$, meaning that $\forall t \in [\delta_{i-1}, \delta_i],\ t \models c_{\bowtie}$. Rule (\ref{restr}), instead, defines the \textit{reset transitions} $(i, s, q) \xrightarrow{\alpha}(1, s', q') \in \martr$ that reset the personal clock $c$ either within the $1^{st}$ Time Region (when $i=1$), or by bringing the agent \textit{back to} the 1$^{st}$ Time Region. In the following, we denote by $\mathcal{R}\subset \martr$ the \textit{set of the reset transitions}.  

To describe the deterministic transitions of $\class_\dta$, instead, we define a set $\mathcal{E}$ of \textit{clock events}. Each clock event has the form $e = (\mathcal{A}_e, \Delta,p_e)$, where $\mathcal{A}_e \subset S_{\dta_i}$ is the \textit{active set}, $\Delta$ is the \textit{duration}, and $p_{e}: \mathcal{A}_e \times S_{\dta} \longrightarrow [0,1]$ is the \textit{probability distribution}. If the agent enters $\mathcal{A}_e$, that is the sets of states in which $e$ can be active, a countdown starts from $\Delta$. When this elapses, $e_i$ is deactivated and the agent is immediately moved to a new state sampled from $p_e((i,s,q),\cdot): S_{\dta} \longrightarrow [0,1]$, where $(i,s,q)\in \mathcal{A}_e$ is the state in which the agent is when the countdown hits zero. Moreover, $e_i$ is deactivated also when the agent takes a reset transition. In $\class_\dta$, we define:
\begin{itemize}
\item  one clock event $e_i\in \dettr$ for each time region $S_{\dta i}$, $i = 2,\ldots, k$;
\item $\ell + 1$ clock events $e_{1}^0, e_1^1, \ldots, e_1^{\ell}\in \dettr$ for the $1^{st}$ Time Region, where $\ell$ is the number of reset events $(1, s, q) \xrightarrow{\alpha}(1, s', q') \in \mathcal{R}$ defined by (\ref{restr}) with $i=1$.
\end{itemize}
For $i>1$, $\mathcal{A}_{i} = S_{\dta i}$, $\Delta_i =\delta_{i}-\delta_{i-1}$, and the probability distribution is 
\begin{equation}\label{pei}
p_{i}((i, s, q), (i', s', q')) = 
\begin{cases} 
1\quad \mbox{if }i' =i+1, s'=s, q'=q,\\
0\quad \mbox{otherwise}.
\end{cases} 
\end{equation}
As it is defined, each clock event $e_i$ with $i>1$ connects each state $(i, s, q) \in \mathcal{A}_i$ with $(i+1, s, q) \in S_{\dta i+1}$, hence, when the duration $\Delta_i$ of $e_i$ elapses, the clock event moves the agent from its state to the equivalent one in the next time region. When $i=1$, instead, the duration and the probability distribution of each clock event $ e_1^j$, $j = 1,\ldots, \ell,$ are defined in the same way as before (i.e. $\Delta_1^j = \delta_{1}-\delta_0 = \delta_1$, and $p_{1}^j$ is given by (\ref{pei})), but the activation sets are now subsets of $S_{\dta_1}$. Indeed, since each reset transition $(1, s, q) \xrightarrow{\alpha_j}(1, s', q') \in \mathcal{R}$ initiates the clock, for each of them, we need to define a clock event $e_1^j$, whose activation set $\mathcal{A}_i^j$ is the set of states in $S_{\dta_1}$ that can be reached by the agent \textit{after} it has taken the reset transition $(1, s, q) \xrightarrow{\alpha_j}(1, s', q')$. Furthermore, we have to define an extra clock event $e_1^0$, with $\mathcal{A}_1^0 = S_{\dta_1}$, $\Delta_1^0 = \delta_{1}$, and $p_{1}^0$ given by (\ref{pei}), that is the only clock event initiated at time $t=0$ (and not by the agent entering $\mathcal{A}_1^0$). Indeed, we require for the \textit{initial state} $s_{0,\dta}$ of $\class_\dta$ to be one of the states of the form $(1,s,q_0)$, where $s \in S$ and $q_0$ is the initial state of $\dta$ (hence, $s_{0,\dta}$ belongs to $\mathcal{A}_1^0$). Finally, since the probability distributions $p_{1}^j$, $\forall j$, are all defined as in (\ref{pei}), also the clock events of the $1^{st}$ Time Region move the agent from a state to the equivalent one in the next time region (the $2^{nd}$), when the countdown from $\Delta_1^j = \delta_1$ elapses. In the following, we denote by $(i, s, q) \dashrightarrow_{e} (i+1, s, q)$ the deterministic transition from $(i, s, q) \in S_{\dta i}$ to $(i+1, s, q) \in S_{\dta i+1}$ encoded by $e \in \dettr$, and by $\boldsymbol\nu_{e,s,q} = \mathbbm{1}_{(i+1,s,q)} - \mathbbm{1}_{(i,s,q)}$ its update vector. The last component of $\class_\dta$ that we define is the \textit{set of final states} $F_\dta$, which is given by $F_\dta = \{(i,s,q)\in S_\dta\ \lvert\ q \in F\}$.

\begin{figure}[t]
\begin{center}
\begin{small}
\begin{tikzpicture}[on grid, shorten <=1pt, >=stealth', auto]
    \node at (-2.57,1.35)  [state, draw=none] (t1) {\begin{tabular}{c}\textit{1$^{st}$ Time Region}\\$0\leq c\leq 5$\end{tabular}};
	\node at (-2.57,-0.9)  [state, draw=none] (t1) {\begin{tabular}{c}\textit{2$^{nd}$ Time Region}\\$c\geq5$\end{tabular}};
	\node at (-1.5,1.2)  [state, draw=none] (i) {};
	\node at (0,1.2)   [state, rectangle,  rounded corners, scale=0.8] (S0)   {$(1,S,q0)$};
    \node at (0,0)   [state, rectangle, rounded corners, scale=0.8] (I0)   {$(1,I,q0)$};
    \node at (0,-0.9)   [state, rectangle, rounded corners, scale=0.8] (S02)   {$(2,S,q0)$};
  	\node at (0,-2.1)   [state, rectangle, rounded corners, scale=0.8] (I02)   {$(2,I,q0)$};
    \node at (2.5,1.2)   [state, rectangle, rounded corners, scale=0.8, color = blue] (S10)   {$(1,S,q1)$};
  	\node at (5,1.2)   [state, rectangle, rounded corners,accepting, scale=0.8, color = blue] (S2)   {$(1,S,q2)$};
  	\node at (5,0)   [state, rectangle, rounded corners,accepting, scale=0.8, color = blue] (I2)   {$(1,I,q2)$};
  	\node at (2.5,-0.9)   [state, rectangle, rounded corners, scale=0.8] (S11)   {$(2,S,q1)$};
  	\node at (2.5,-2.1)   [state, rectangle, rounded corners, scale=0.8] (I11)   {$(2,I,q1)$};
  	\node at (5,-0.9)   [state, rectangle, rounded corners,accepting, scale=0.8] (S21)   {$(2,S,q2)$};
  	\node at (5,-2.1)   [state, rectangle, rounded corners,accepting, scale=0.8] (I21)   {$(2,I,q2)$};
    \node at (3.7,0.0)  [state, draw=none] (q) {};
    \node at (3.7,-2.1)  [state, draw=none] (q1) {};
    \node at (4.15,0.4)  [state, draw=none] (q2) {};
    \node at (4.15,-2.5)  [state, draw=none] (q3) {};
    \node at (6.3,1.2)  [state, draw=none] (q7) {};
    \node at (6.3,-0.9)  [state, draw=none] (q17) {};
    \node at (5.85,1.6)  [state, draw=none] (q27) {};
    \node at (5.85,-1.3)  [state, draw=none] (q37) {};
    \node at (-1.3,0.0)  [state, draw=none] (0q) {};
    \node at (-1.3,-2.1)  [state, draw=none] (0q1) {};
    \node at (-0.85,0.4)  [state, draw=none] (0q2) {};
    \node at (-0.85,-2.5)  [state, draw=none] (0q3) {};
    \node at (1.3,1.2)  [state, draw=none] (0q7) {};
    \node at (1.3,-0.9)  [state, draw=none] (0q17) {};
    \node at (0.85,1.6)  [state, draw=none] (0q27) {};
    \node at (0.85,-1.3)  [state, draw=none] (0q37) {};
 	\node at (5.1,-0.86)  [state, draw=none] (s12) {};
    \node at (4.9,-0.82)  [state, draw=none] (s11) {};
    \node at (5.1,-2.18)  [state, draw=none] (i12) {};
    \node at (4.9,-2.15)  [state, draw=none] (i11) {};
    \node at (5.1,1.24)  [state, draw=none] (s22) {};
    \node at (4.9,1.28)  [state, draw=none] (s21) {};
    \node at (5.1,-0.08)  [state, draw=none] (i22) {};
    \node at (4.9,-0.05)  [state, draw=none] (i21) {};

  	\path [->]	 (S0)		edge [left]	node {$inf$}(I0)
  				 (I0)		edge [above, color=red]	node {$rec$}(S10)
  				 (S10)		edge [below, near start]	node {$inf\ \ \ $}(I2)
  				 (S11)		edge [right]	node {$inf$}(I11)
  				 (I11)      edge [right, near end, bend left, color=red, out = 50, in = 130]	node {$rec$}(S10)
  				 (s11)		edge [left]	node {$inf$}(i11)
  				 (i12)		edge [right]	node {$rec$}(s12)
				 (s21)		edge [left]	node {$inf$}(i21)
  				 (i22)		edge [right]	node {$rec$}(s22)
  				 (S02)		edge [left]	node {$inf$}(I02)
  				 (I02)		edge [right, very near start, color=red, out = 35, in = 213]	node {$rec$}(S10)
 				 (S10)		edge [dashed, color = blue] 	node {}(S11)
  				 (q1)		edge [dashed, color = blue]			    node {}(I21)
  				 (q17)		edge [dashed, color = blue]			    node {}(S21)
  				 (0q1)		edge [dashed, color = green!50!black]			    node {}(I02)
  				 (0q17)		edge [dashed, color = green!50!black]			    node {}(S02)
  				 (i)		edge []			    node {}(S0)
  						
  		  [-]    (q)		edge [dashed, color = blue]			    node {}(I2)
  			     (q2)		edge [dashed, color = blue]			    node {}   (q3)
  			     (q7)	    edge [dashed, color = blue]			    node {}(S2)
  				 (q27)		edge [dashed, color = blue]			    node {}   (q37)
  				 (0q)		edge [dashed, color = green!50!black]			    node {}(I0)
  				 (0q2)		edge [dashed, color = green!50!black]			    node {}   (0q3)
  				 (0q7)		edge [dashed, color = green!50!black]			    node {}(S0)
  				 (0q27)		edge [dashed, color = green!50!black]			    node {}   (0q37);
	
	\begin{scope}[on background layer]
	\node at (1.05,0.4) [state, rectangle, draw=white, fill=black!5,minimum width =30.5em,minimum height =8.6em] {};
	\node at (1.05,-1.5) [state, rectangle, draw=white, fill=black!10,minimum width =30.5em,minimum height =6.6em] {};
	\end{scope}
\end{tikzpicture}
\end{small}
\end{center}
\caption[]{The agent class $\class_{\dta}$ associated with the DTA $\dta$ of the running example.}
\label{Fig:agentclass}
\end{figure}
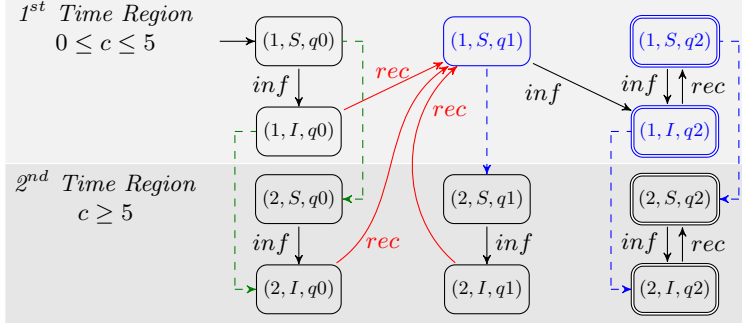

\paragraph{Example.} Fig. $\ref{Fig:agentclass}$ represents the product $\class_\dta$ between the agent class $\class$ and the property $\dta$ of the running example (Fig. \ref{fig:sis}). The state $(1,I,q1)$ that cannot be reached by the single agent is omitted. The black transitions are the Markovian transitions without reset; the \textcolor{red}{red} transitions are the Markovian transitions that reset the clock; finally, we define 2 clock events, \textcolor{OliveGreen}{$e^0_1$} and \textcolor{blue}{$e^1_1$},  with duration $\Delta = 5$ for the $1^{st}$ Time Region, and the dashed \textcolor{OliveGreen}{green} (resp. \textcolor{blue}{blue}) transitions are the deterministic transitions encoded by \textcolor{OliveGreen}{$e^0_1$} (resp. \textcolor{blue}{$e^1_1$}). In \textcolor{blue}{blue}, we also highlight the states that belong to the activation set \textcolor{blue}{$\mathcal{A}_{e_1^1}$} (while \textcolor{OliveGreen}{$\mathcal{A}_{e_1^0}$} is the whole $1^{st}$ Time Region). Intuitively, the agent can be found in one of the states belonging to the $1^{st}$ Time Region whenever its personal clock $c$ is between 0 and 5, i.e. less that 5 time units have passed since $t=0$ or since a recovery \textcolor{red}{$rec$}. In a similar way, the agent is in the 2$^{nd}$ Time Region when the valuation of $c$ is above $5$. Moreover, when the the duration of the clock events elapses (i.e. the countdown from 5 hits 0), the agent is moved from the 1$^{st}$ Time Region to the 2$^{nd}$ Time Region by the deterministic \textcolor{OliveGreen}{green} and \textcolor{blue}{blue} transitions, that indeed have duration $\Delta= 5$ and are initiated at $t=0$ or by the reset actions \textcolor{red}{$rec$}, respectively. At the end, the agent is in one of the final states ($(1,S,q2),\ (1,I,q2),\ (2,S,q2)$ or $(2,I,q2)$) at time $T$, if it meets property $\dta$ within time $T$, i.e. within $T$, the agent has been infected during the 5 time units that follow a recovery. Hence, to verify $\dta$, we will compute the probability of being in one of the final states of $\class_\dta$ at time $T$.

\mypar{The IMRP $Z_{\class_\dta}(t)$ and the Satisfaction Probability $P(T)$} Now we show how to formally define the IMRP $Z_{\class_\dta}(t)$ that describes the state of the product $\class_\dta$ in the mean field regime. In particular, we derive the \textit{Delay Differential Equations} (DDE) \cite{cinlar} for the \textit{transient probability} $\boldsymbol P(t)$ of $Z_{\class_\dta}(t)$, in terms of which we compute the \textit{satisfaction probability} $P(T)$.

Let $\fluid(t)$ be the Fluid Approximation of the population model $\pop\upsize$. To define the transient probability $\boldsymbol P(t)$ of $Z_{\class_\dta}(t)$, we exploit the fact that, in the case of an IMRP, we have: $\frac{d\boldsymbol P}{dt}(t) = \boldsymbol{M}(\fluid(t))\boldsymbol P(t) + \boldsymbol{D}(\fluid(t), \boldsymbol P(t))$ (cf. \cite{cinlar}). In this equation, $\boldsymbol{M}(\fluid(t))$ is the \textit{generator matrix} for the Markovian transitions, and $\boldsymbol{D}(\fluid(t),\boldsymbol P(t))$ accounts for the deterministic events. The elements of $\boldsymbol{M}(\fluid(t))$ are computed following the same procedure that was described in Sec. \ref{subsec:fastsim}, where the multiplicity of each transition $(i, s, q) \xrightarrow{\alpha}(i, s', q') \in \martr$ in $\class_\dta$ is always equal to 1 (one single agent) and the Lipschitz limit $f_{\alpha}(\fluid(t))$ of $\alpha$ is that of the rate of the transition $s\xrightarrow{\alpha}s'$ in $\pop\upsize$ from which $\alpha$ was derived (by rules (\ref{martr}) or (\ref{restr})). 

To define the components of $ \boldsymbol{D}(\fluid(t), \boldsymbol P(t))$, instead, consider any clock event $e  = (\mathcal{A}_i, \Delta_i, p_i) \in \dettr$, \textit{except }$e_1^0$, whose contribute will be computed later on\footnote{If $e$ is one of events of the $1^{st}$ Time Region, i.e. $e = e_1^j$, for some $j=1,\ldots,\ell,$ in this section, we drop the index $j$ to ease the notation, i.e. we write $e_1^j = e_1 = (\mathcal{A}_1, \Delta_1, p_1)$.}. Choose one of the deterministic transitions $(i, s, q) \dashrightarrow_{e_i} (i+1, s, q)$  encoded by $e_i$.  The agent takes this transition at time $t$ when: (1) it entered $\mathcal{A}_i \subseteq S_{\dta_i}$ at time $t-\Delta_i$ (initiating its personal clock), and (2) it is in state $(i, s, q)\in \mathcal{A}_i$ at time $t$ (when the duration of $e_i$ elapses). Hence, to compute the term that corresponds to this transition in $\boldsymbol{D}(\fluid(t), \boldsymbol P(t))$, we need to: (1) record the flux of probability that entered $\mathcal{A}_i$ at time $t-\Delta_i$, and (2) multiply it by the probability that the agent reaches $(i, s, q)\in \mathcal{A}_i$ at time $t$, conditional on the state at which it entered $\mathcal{A}_i$ at $t-\Delta_i$.

To compute the probability of step (2), we need to keep track of the dynamics of the agent while the clock event $e_i$ is active. For this purpose, let $\bar{\mathcal{A}_i}$ be the activation set $\mathcal{A}_i$ of $e_i$ extended to contain an extra state $s_{out} = (i, s_{out}, q_{out})$, and let $\bar{\mathcal{M}}$ be the set $\mathcal{M}$ of Markovian transitions in $\class_\dta$ modified in order to make the reset transitions that start in $\mathcal{A}_i$ finish into $s_{out}$ (i.e. for every $(i, s, q) \xrightarrow{\alpha}(i', s', q') \in \mathcal{R}\subset \mathcal{M}$, we define $(i, s, q) \xrightarrow{\alpha}s_{out} \in \bar{\mathcal{M}}$), and to have $s_{out}$ absorbing\footnote{The absorbing state $s_{out}$ is needed for the probability $\boldsymbol{Y}_i(t)$ of step (2) to be well defined. Indeed, the agent can deactivate $e_i$ by taking a reset transition.}. Let $\gen_i(\fluid(t))\in \mbox{Matr}(\lvert\bar{\mathcal{A}_i}\lvert \times \lvert\bar{\mathcal{A}_i}\lvert)$ be the time-dependent matrix s.t.
\begin{equation}\label{gen}
\left(\gen_i(\fluid(t))\right)_{(i,s,q),(i,s',q')} = \sum_{(i, s, q) \xrightarrow{\alpha}(i, s', q') \in \bar{\martr}}\left[\frac{1}{\Phi_{s}(t)}f_{\alpha}(\fluid(t))\right], 
\end{equation}
where again the Lipschitz limit $f_{\alpha}(t)$ of each $\alpha \in \bar{\martr}$ is that of the transition $s\xrightarrow{\alpha}s'$ in $\pop\upsize$ from which its copy $\alpha\in\martr$ was derived (by (\ref{martr}) and (\ref{restr})). Moreover, let the diagonal elements of $\gen_i(\fluid(t))$ to be defined so that the rows sum up to zero. Then, we introduce the \textit{probability matrix} $\boldsymbol{Y}_i(t)$, which is computed in terms of the \textit{generator} $\gen_i(\fluid(t))$ according to the following ODEs (see also \cite{FMC}):
\begin{equation}\label{probability}
\begin{cases}
\frac{d\boldsymbol{Y}_i}{dt}(t) = \boldsymbol{Y}_i(t)\gen_i(\fluid(t))- \gen(\fluid(t-\Delta_i))\boldsymbol{Y}_i(t),\quad\ \Delta_i \leq t \leq T,  \\
\frac{d\boldsymbol{Y}_i}{dt}(t) = \boldsymbol{Y}_i(t)\gen_i(\fluid(t)),\quad\ 0 \leq t \leq \Delta_i,\\
\end{cases}
\end{equation}
with $\boldsymbol{Y}_i(0) = \mbox{\textbf{I}}$. By definition, $(\boldsymbol{Y}_i(t))_{(i,s',q'),(i,s,q)}$ is the Fluid Approximation of the probability of step (2), i.e. the probability that the agent, which has entered $\mathcal{A}_i$ in state $(i,s',q')$ at time $t-\Delta_i$, moves (Markovianly) within $\mathcal{A}_i$ for $\Delta_i$ units of time, and reaches $(i,s,q) \in \mathcal{A}_i$ at time $t$ (exactly when $e_i$ elapses).

In terms of the probability matrix $\boldsymbol{Y}_i(t)$, we can now define the component of $\boldsymbol{D}(\fluid(t),\boldsymbol P(t))$ that corresponds to the deterministic transition $(i, s, q) \dashrightarrow_{e_i} (i+1, s, q)$ of the clock event $e_i \in \dettr$. This component is the element in position $((i, s, q),(i+1, s, q))$ in $\boldsymbol{D}(\fluid(t),\boldsymbol P(t))$, we call it $D_{e_i, s,q}(\fluid(t), \boldsymbol P(t))$, and is \mbox{given by}

\vspace{-5mm}
\begin{align*}
D_{e_i, s,q}(\fluid(t),\boldsymbol P(t))=& \hspace{-1mm}\sum_{(i,\bar{s},\bar{q})\in \mathcal{A}_{i}} \left[ \mbox{\textbf{1}}_{\{i>1\}} D_{e_{i-1}, \bar{s},\bar{q}}(\fluid(t-\Delta_{i}),\boldsymbol P(t-\Delta_i))+\mbox{\textbf{1}}_{\{i=1\}}\times\phantom{\sum_{(i,s,q)\in S_{\dta_i}}}\right.  \\
\times\sum_{(i', s', q') \xrightarrow{\alpha}(1, \bar{s}, \bar{q})\in \mathcal{R}}&\left.\phantom{\sum_{(i,s,q)\in S_{\dta_i}}}\hspace{-15mm}\frac{1}{\Phi_{s'}(t)} f_{\alpha}(\fluid(t-\Delta_1))(\boldsymbol P(t-\Delta_1))_{(i',s',q')}\right]\hspace{-1.5mm}(\boldsymbol{Y}_i(t))_{(i,\bar{s},\bar{q}),(i,s,q)},\numberthis\label{ddd}
\end{align*}

\vspace{-3mm}
\noindent where $(\boldsymbol P(t-\Delta_1))_{(i',s',q')}$ is the component in position $(i',s',q')\in S_{\dta_{i'}}$ in the vector of the transient probability $\boldsymbol P(t-\Delta_1)$ of $Z_{\class_\dta}$ at time $t-\Delta_1$. In (\ref{ddd}), for each state $(i,\bar{s},\bar{q})$ in the activation set $\mathcal{A}_i$, the quantity inside the squared brackets is the probability flux that entered $(i,\bar{s},\bar{q})$ at time $t-\Delta_i$. In particular, when $i>1$, $D_{e_{i-1}, \bar{s},\bar{q}}(\fluid(t-\Delta_{i}),\boldsymbol P(t-\Delta_i))$ accounts for the termination of clock event $e_{i-1}$ (i.e. the deterministic transition $(i-1, \bar s,\bar q) \dashrightarrow_{e_i} (i, \bar s, \bar q)$ fired at time $t-\Delta_i$). When we consider the $1^{st}$ Time Region, i.e. $i=1$, instead, each term in the sum over the reset transitions is the flux of probability entering $(1,\bar{s}, \bar{q})$ at time $t-\Delta_1$ due to a clock reset. Finally, $(\boldsymbol{Y}_i(t))_{(i,\bar{s},\bar{q}),(i,s,q)}$ is again the probability of reaching $(i,s,q)\in \mathcal{A}_i$ from $(i,\bar{s},\bar{q})\in \mathcal{A}_i$ in $\Delta_i$ units of time. 

All the other off-diagonal elements of $\boldsymbol D(\fluid(t),\boldsymbol P(t))$ can be computed in a similar way, while the diagonal ones are defined so that the rows sum up to zero. Moreover, since at the end $\boldsymbol D(\fluid(t),\boldsymbol P(t))$ depends on the state of the system at times $t-\Delta_1, \ldots, t-\Delta_k$ (through the probabilities $\boldsymbol Y_i(t),\ i = 1,\ldots, k$), we write $\boldsymbol D(\fluid(t)) = \boldsymbol D(\fluid, \boldsymbol P, \Delta_1, \ldots, \Delta_k,t)$. Then, we define the \textit{transient probability} $\boldsymbol P(t)$ of the IMRP $Z_{\class_\dta}(t)$ as the solution of the following system of DDEs:

\vspace{-4mm}
\begin{align*}
\boldsymbol{P}(t)= &\hspace{-1mm}\int_{0}^{t}\hspace{-1mm}\boldsymbol M(s)\boldsymbol P(t)ds +\hspace{-1mm} \int_{0}^t\hspace{-1mm}\boldsymbol D(\fluid,\boldsymbol P, \Delta_1, \ldots, \Delta_k,s)ds + \boldsymbol 1_{t\geq\Delta_{1}}\hspace{-2.5mm} \sum_{(s,q)\in S\times Q}\hspace{-1mm}y_{e_1^0}\boldsymbol \nu_{e_1^0, s, q}.\numberthis\label{fmc}
\end{align*}

\vspace{-3mm}
\noindent In (\ref{fmc}), the third term is a deterministic jump in the value of $\boldsymbol{P}(t)$ at time $t = \Delta_1$, and represents the contribute of the clock event $e_1^0$. In such term, the vectors $\boldsymbol \nu_{e_1^0, s, q}$ are the update vectors of the deterministic transitions encoded by $e_0^1$  (hence, the sum is computed over all such transitions), and the probability $y_{e_1^0}$ is the value at time $t=\Delta_1$ of the component in position ${(s_{0,\dta},(1,s,q))}$ (where $s_{0,\dta}$ is the initial state of $\class_\dta$) in the matrix $\boldsymbol{Y}_{e_1^0}(t)$ defined by:

\vspace{-2mm}
$$
\frac{d\boldsymbol{Y}_{e_1^0}}{dt}(t) = \boldsymbol{Y}_{e_1^0}(t)\gen_1(\fluid(t)),\quad\ 0 \leq t \leq \Delta_1,
$$
with $\gen_1(\fluid(t))$ defined as in (\ref{gen}), and $\boldsymbol{Y}_{e_1^0}(0)= \boldsymbol I$. Hence, $y_{e_1^0}\hspace{-1mm}=\hspace{-1mm}(\boldsymbol{Y}_{e_1^0}(\Delta_1))_{s_{0,\dta},(1,s,q)}$ is the probability that, starting form  $s_{0,\dta}$, the agents reaches $(1,s,q)\in S_{\dta_1}$ at time $t=\Delta_1$ (exactly when the deterministic event $(1,s,q)\dasharrow_{e_1^0}(2,s,q)$ fires).
  
Given the product $\class_\dta$, the IMRP $Z_{\class_\dta}(t)$, and its transient probability $\boldsymbol P(t)$, the following result holds true.
\begin{proposition}\label{finalprop}
There is a 1:1 correspondence between $\Sigma_{\class, \dta, T}$ and the set $AccPath(\class_\dta, T)$ of accepted paths of duration $T$ of $\class_\dta$. Hence, 
$$
P(T)= Prob_Z \{\Sigma_{\class, \dta, T}\} = Prob_{Z_{\class_\dta}}\{AccPath(\class_\dta, T)\} = P_{F_\dta}(T),
$$
where $Prob_{Z_{\class_\dta}}$ is the probability measure defined by $Z_{\class_\dta}$, and $P_{F_\dta}(T)$ is the sum of the components of $\boldsymbol P(T)$ corresponding to the final states $F_\dta$ of $\class_\dta.\qquad\qquad\qed$
\end{proposition}
In other words, according to Proposition \ref{finalprop}, when the population of $\pop\upsize$ is large enough, $P_{F_\dta}(T)$ is an accurate approximation of the probability that a (random) single agent in $\pop\upsize$ satisfies property $\dta$ within time $T$.

 
\vspace{1mm}
\noindent\textit{Example.} For the product $\class_\dta$ in Fig. \ref{Fig:agentclass}, the non-zero off-diagonal entries of the generator matrix $\boldsymbol G_{e_1^1}(\fluid(t))$ of the clock event $e_1^1$ are: $G_{(S,q1)(I,q2)}(t) = k_i\Phi_{I}(t); \ G_{(S,q2)(I,q2)}(t) = k_i\Phi_{I}(t);$ and $G_{(I,q2)(S,q2)}(t) = k_r$. In terms of $\boldsymbol G_{e_1^1}(\fluid(t))$, we can define $\boldsymbol Y_{e_1^1}(t)$, as in (\ref{probability}), that is then used in the DDEs (\ref{fmc}) for the probability $\boldsymbol{P}(t)$. In this latter set of 9 DDEs (one for each state of $\class_\dta$), we have:

\vspace{-2mm}
\begin{small}
\begin{align*}
\boldsymbol{P}_{(1,S,q1)}(t) =&\ \textcolor{red}{\int_{0}^{t}k_r \boldsymbol{P}_{(1,S,q1)}(s)ds}\ \textcolor{black}{-\int_{0}^{t}k_i\Phi_{I}(s)\boldsymbol{P}_{(1,S,q1)}(s) ds}\ +\\
&\textcolor{blue}{-\ \int_{0}^{t}k_r\boldsymbol Y_{(1,S,q1),(1,S,q1)}(s-5,s)\boldsymbol{P}_{(1,S,q1)}(s)ds}.
\end{align*}
\end{small}

\vspace{-2mm}
\noindent\textit{Remark.} The presence of \textit{only one clock} in $\dta$ enables us to define $\class_\dta$ in such a way that $Z_{\class_\dta}(t)$ is an IMRP. This cannot be done when we consider \textit{multiple clocks} in $\dta$. Indeed, in the latter case, the definition of the stochastic process which describes the state of the product $\class_\dta$ is much more complicated, since, when a reset event occurs, we still need to keep track of the valuations of all the other clocks in the model (hence, the dynamics between the time regions of $\class_\dta$ is not as simple as in the case of one single clock). In the future, we plan to investigate possible extensions of our model checking procedure to timed properties with multiple clocks, also taking into account the results of \cite{fu} and \cite{chen}.

\vspace{-2mm}
\subsection{The Mean Behaviour of the Population Model}
\label{sec:Fluid}

It is possible to modify our FMC procedure in order to compute the \textit{mean} number of agents that satisfy $\dta$. This can be done by assigning a personal clock to each agent, and monitoring all of them using as agent class the product $\class_\dta$ defined in Sec. \ref{sec:FMC}. In terms of $\class_\dta$, we can build the population model  $\pop_\dta$, with $\class_\dta$ as the only agent class, and the sum $P_{F_{\dta}}(T)$ of the components corresponding to the final states of $\class_\dta$ in the Fluid Approximation $\fluid(t)$ of $\pop_\dta$ computed at $t=T$ is indeed the mean number of agents satisfying property $\dta$ within time $T$. The construction of $\pop_\dta$ is not difficult: it follows the procedure of \cite{qest}, where a little extra care has to be taken just in the definition of the global transitions of $\pop_\dta$. Indeed, if we build for instance the population model $\pop_\dta$ of the running example, we need to consider that the infected individual that passes the virus to an agent in state $(1,S,q0)$ can be now in one of \textit{five} states: $(1,I,q0), (1,I,q2), (2,I,q0), (2,I, q1)$ or $(2,I,q2)$. For this reason, we have to define \textit{five} Markovian global transition in $\pop_\dta$, each of which moves an agent from $(1,S,q0)$ to $(1,I,q0)$ at a rate that is influenced by the number of individuals that are in the infected states of $\class_\dta$, recorded in the counting variables $X_{(1,I,q0)}(t), X_{(1,I,q2)}(t), X_{(2,I,q0)}(t), X_{(2,I, q1)}(t)$ or $X_{(2,I,q2)}(t)$. The same reasoning has to be followed for the definition of the infections of the agents in states $(1,S,q1), (1,S,q2), (2,S,q0), (2,S,q1)$ and $(2,S,q2)$. At the end, as for the single agent, due to the deterministic events, the Fluid Approximation $\fluid(t)$ of $\pop_\dta$ is the solution of a system of DDEs similar to (\ref{fmc}). The definition of such approximating equations for a population model with exponential and deterministic transitions is not new \cite{hayden}, but, even if the results are promising (see Sec. \ref{sec:ExpRes}), to our knowledge, nobody has yet proven the convergence of the estimation in the limit $N\rightarrow+\infty$. We save the investigation of this result for future work.

\begin{small}
\begin{table}[t]
\begin{center}
Fluid Model Checking\\
\begin{tabular}{|c||c|c|c||c|c|c|}
\hline
    $N$    &\phantom{.}MeanRelErr\phantom{.}&\phantom{.}MaxRelErr\phantom{.}&\phantom{.}RelErr(T)\phantom{.}&\phantom{.}TimeDES\phantom{.}&\phantom{.}TimeFMC\phantom{.}  &\phantom{.} Speedup\phantom{.}\\ 
    \hline
    250    &  0.0927    &  6.4512   &  0.1043   &  11.0273  &  0.4731  &  23.3086\\
    \hline
    500    &  0.0204    &  1.7191   &  0.0048   &  44.0631  &  0.3980  &  110.7113\\
    \hline
    1000   &  0.0118    &  0.7846   &  0.0003   & 170.9154  &  0.3998  &  427.5022\\
    \hline
\end{tabular}\\

\vspace{1mm}
Fluid Approximation of the mean behaviour\\
\begin{tabular}{|c||c|c|c||c|c|c|}
\hline
    $N$    &\phantom{.}MeanRelErr\phantom{.}&\phantom{.}MaxRelErr\phantom{.}&\phantom{.}RelErr(T)\phantom{.}&\phantom{.}TimeDES\phantom{.}&\phantom{.}TimeFluid\phantom{.}  &\phantom{.} Speedup\phantom{.}\\
    \hline
    250    &  0.1127    &  0.2316    &  0.0921   & 105.5647  &  0.4432  &  339.7217\\
    \hline
    500    &  0.0289    &  0.3177    &  0.0289   & 415.0635  &  0.4237  &  979.6165\\
    \hline
    1000   &  0.0117    &  0.2216    &  0.0117   & 1547.0340 &  0.4213  & 3672.0484\\
    \hline
\end{tabular}
\end{center}

\caption{Mean Relative Error (MeanRelErr), Maximum Relative Error (MaxRelErr), and Relative Error at final time (RelErr(T)) of the FMC (top) and the Fluid Approximation of the mean behaviour (bottom) for different values of $N$. The table enlists also the execution times (in seconds) of the DES (TimeDES) and the approximations (TimeFMC and TimeFluid), and the speedups (TimeDES divided by the other times).}
\label{table:error}
\end{table}
\end{small}



\section{Experimental Results}
\label{sec:ExpRes}

\begin{figure}[t!]
\begin{center}
\begin{center}
\includegraphics[width=.4\textwidth]{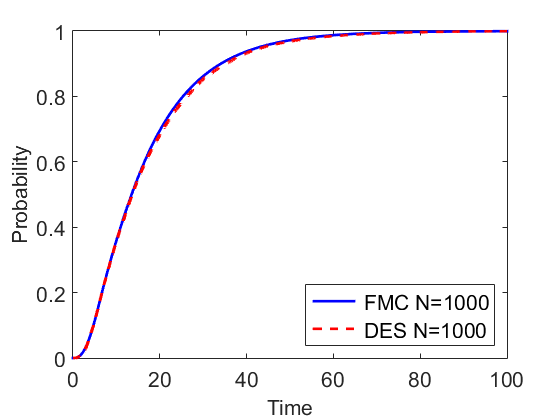}
\includegraphics[width=.4\textwidth]{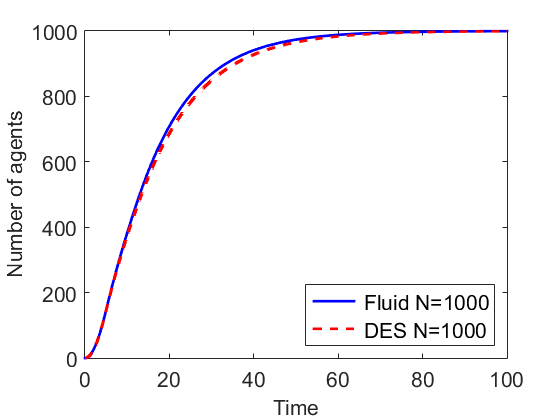}
\end{center}
\end{center}
\caption{The satisfaction probability $P(T) = P_{F_\dta}(T)$ obtained by the Fluid Model Checking (left) and the Fluid Approximation of the mean behaviour (right) in the case $N=1000$. The results are compared with those obtained by the DES.}
\label{fig:results}
\end{figure}

To validate the procedures of Sec. \ref{sec:FMC}, we performed a set of experiments on the running example, where we fixed: $k_i = 1.2$, $k_r = 1$, $\Delta = 5$, and an initial state of the population model with a susceptible-infected ratio of 9:1. As in Fig. \ref{Fig:agentclass}, we let the single agent start in the susceptible state, and we considered three different values of the population size: $N=250, 500,1000$. For each $N$, we compared our procedures with a statistical estimate from 10000 runs, obtained by a dedicated Java implementation of a Discrete Event Simulator (DES). The errors and the execution times obtained by our FMC procedure (top) and the Fluid Approximation of the mean behaviour (bottom) are reported in Tab. \ref{table:error}. Regarding the errors, we would like to remark that the Relative Errors (RE) of both the FMC and the Fluid Approximation reach their maximum in the very first instants of time, when the true satisfaction probability (i.e. the denominator of the REs) is indeed really small, but then they decay really fast as the values of $P_{F_\dta}(t)$ increase (this can be easily deduced from the values of the mean REs and the REs at final time). As expected, the accuracy of the approximations increases with the population size, and is already reasonably good for $N=500$. Moreover, the resolution of the DDEs is computationally independent of $N$, and also much faster (approximatively 3 orders of magnitude in the case of the Fluid for $N = 1000$) than the simulation based method. Fig. \ref{fig:results} shows the results of the FMC and the Fluid Approximation in the case N=1000.
\vspace{-2mm}

 
 \section{Conclusions}
 \label{sec:conc}
 
We defined a fast and efficient FMC procedure that accurately estimates the probability that a single agent inside a large collective system satisfies a time-bounded property specified by a single-clock DTA. The method requires the integration of a system of DDEs for the transient probability of an IMRP, and the exactness of the estimation is guaranteed in the limit of an infinite population.

\vspace{1mm}
\noindent\textit{Future Work.} During the experimental analysis, we realised that, on certain models and properties, the DDEs (\ref{probability}) can be \textit{stiff}, and their numerical integration in MATLAB is unstable (see also \cite{FMC}). In the future, we want to address this issue by considering alternative integration methods \cite{fortran}, investigating also numerical techniques for MRP with time-dependent rates\cite{timenet}. Furthermore, we plan to prove the convergence of the Fluid Approximation of Sec. \ref{sec:Fluid}, and to investigate higher-order estimates as in \cite{qest, epew}. Finally, we want to extend the FMC procedure of this paper to validate requirements specified in the logic CSL$^{TA}$ \cite{cslta} and DTA properties endowed with multiple clocks (possibly considering the approximation techniques defined in \cite{fu} and \cite{chen}).
\vspace{-1mm}




\bibliographystyle{plain}
\bibliography{biblio}


\clearpage

\appendix

\section{Proofs}
\label{appendix}

\subsection{$Prob_{Z\upsize}$ and $Prob_{Z}$, and Measurability of $\Sigma_{\class, \dta, T}$}
\label{app:meas}

Consider a single agent of class $\class = (\cstsp,\ctrsp)$ in a population model  $\pop\upsize = (\class, \ptrsp\upsize, \pinst\upsize)$, and a timed property $\dta = \dta(T) = (T, \lbset, \atprop,  \mathcal{C}\mathcal{C}, \gstsp,\ginst, \gfinst, \gtr)$. Let $\Sigma_{\class, \dta, T}$ be the set of time-bounded paths of $\class$ accepted by $\dta$ introduced in Sec. \ref{subsec:properties}, and let $Z\upsize(t)$ and $Z(t)$ be the two stochastic processes defined for the Fast Simulation in Sec. \ref{subsec:fastsim}.

The process $Z\upsize(t)$ is non-Markovian, and it becomes a CTMC only when considered in couple with the state $\boldsymbol X\upsize(t)$ of $\pop\upsize$. Indeed, given the Markov process $\boldsymbol Y = (Y_1\upsize(t), \ldots, Y_N\upsize(t))$, that enlists all the stochastic processes that describe the state of the $N$ agents in $\pop\upsize$, $Z\upsize(t)$ is defined to be the projection of $\boldsymbol Y$ on the component $Y_i\upsize(t)$ that represents the single agent that we are considering in our FMC procedure. For this reason, if we assume w.l.o.g. that $i=1$, i.e. the stochastic process for the single agent is the first component of $\boldsymbol Y$, the transient probabilities of $Z\upsize(t)$ and $(Y_1\upsize(t), \ldots, Y_N\upsize(t))$ are such that
\begin{equation}\label{tr}
\boldsymbol P\left\lbrace Z\upsize(t) = k\right\rbrace = \sum_{\boldsymbol s\in S^{N-1}}\boldsymbol P\left\lbrace(Y_1\upsize(t), \ldots, Y_N\upsize(t)) = (k,\boldsymbol s)\right\rbrace.
\end{equation}
Then, exploiting the equality (\ref{tr}), the \textit{probability measure} $Prob_{Z\upsize}$ over path of $Z\upsize$ can be easily derived form $Prob_{(Z\upsize, \boldsymbol X\upsize)}$, which is the probability measure of the CTMC $(Z\upsize, \boldsymbol X\upsize)$ defined in the standard way over cylinder sets of paths (cf e.g. \cite{katoen}).

The \textit{probability measure} $Prob_{Z}$ over the paths of the stochastic process $Z(t)$, instead, is the standard probability measure for ICTMC (cf. e.g. \cite{FMC}).

Then, the measurability of $\Sigma_{\class, \dta, T}$ for $Prob_{Z\upsize}$ and $Prob_{Z}$ is just an adaptation of Theorem 3.2 of \cite{katoen} to $Z\upsize(t)$ and $Z(t)$.

\subsection{Convergence of the Approximation}
\label{app:conv}

Consider a single agent of class $\class = (\cstsp,\ctrsp)$ in a population model  $\pop\upsize = (\class, \ptrsp\upsize, \pinst\upsize)$, and a timed property $\dta = \dta(T) = (T, \lbset, \atprop,  \mathcal{C}\mathcal{C}, \gstsp,\ginst, \gfinst, \gtr)$. Let $\Sigma_{\class, \dta, T}$ be the set of time-bounded paths of $\class$ accepted by $\dta$ introduced in Sec. \ref{subsec:properties}, and let $Z\upsize(t)$ and $Z(t)$ be the two stochastic processes defined for the Fast Simulation in Sec. \ref{subsec:fastsim}. Consider the probability measures $Prob_{Z\upsize}$ and $Prob_{Z}$ of $Z\upsize(t)$ and $Z(t)$, respectively, and define the satisfaction probabilities $P\upsize(T) = Prob_{Z\upsize}\left\lbrace\Sigma_{\class, \dta, T}\right\rbrace$ and $P(T) = Prob_{Z}\left\lbrace\Sigma_{\class, \dta, T}\right\rbrace$. Then the following theorem holds true.

\noindent\textbf{Theorem \ref{th:convergence}}
\textit{For any  $T < +\infty,$}
$$
\lim_{N\rightarrow+\infty}P\upsize(T) = P(T).
$$

\medskip

\begin{proof}
By a standard coupling argument, we can assume that both ${Z\upsize}$ and ${Z}$ are defined on the same sample space $\Omega$. Moreover, for each  $\omega \in \Omega$, let $Z\upsize(\omega, t)$ and $Z(\omega, t)$ denote the state reached at time $t$ in the path corresponding to $\omega$ of $Z\upsize(t)$ and $Z(t)$, respectively. Then, according to Theorem \ref{FastSimTheorem}, for each time horizon $T<+\infty$, there exist a sequence $\epsilon_N\in \mathbb{R}^+$, $\epsilon_N \xrightarrow{N\rightarrow +\infty} 0$, such that
$$
Prob\left\lbrace \omega \in \Omega\ \lvert\ \forall t\leq T,\ Z\upsize(\omega, t)=Z(\omega, t)\right\rbrace \geq 1-\epsilon_N.
$$

Now, fix $T$ and $N$, let $\Sigma_T$ be the set of paths of total duration $T$ for an agent of class $\class$, and consider the (measurable) function $\chi_\dta: \Sigma_T \rightarrow \{0,1\}$, whose value $\chi_\dta(\sigma)$ is $1$ if the path $\sigma\in\Sigma_T$ is accepted by $\dta$ (i.e. the labels of $\sigma$ define a path of $\dta$ that ends in one of the final states $F$), and $0$ otherwise. By definition, $P\upsize(T) = \mathbb{E}[\chi_\dta(Z\upsize)]$ and  $P(T) = \mathbb{E}[\chi_\dta(Z)]$, where $\chi_\dta$ is evaluated on the paths of $Z\upsize$ and $Z$ restricted up to time $T$. Moreover, define 
$$
\Omega_1 = \left\lbrace\omega \in \Omega\ \lvert\ Z\upsize(\omega, t)=Z(\omega, t),\ \forall t\in[0,T]\right\rbrace,
$$ 
and $\Omega_0 = \Omega\setminus\Omega_1$. Then, $\chi_\dta(Z\upsize) = \chi_\dta(Z)$ on $\Omega_1$ and $Prob(\Omega_0) \leq \epsilon_N$. Hence, we have
\begin{align*}
\left|\hspace{-0.5mm}\left|\mathbb{E}\left[\chi_\dta(Z\upsize)\right]-\mathbb{E}\left[\chi_\dta(Z)\phantom{Z\upsize}\hspace{-7.4mm}\right]\right|\hspace{-0.5mm}\right|&\leq\mathbb{E}\left[\left|\hspace{-0.5mm}\left|\ \chi_\dta(Z\upsize) - \chi_\dta(Z)\ \right|\hspace{-0.5mm}\right|\right]=\\
&=\int_{\Omega_1} \left|\hspace{-0.5mm}\left|\chi_\dta(Z\upsize) - \chi_\dta(Z)\right|\hspace{-0.5mm}\right| d\mu_{\Omega}\hspace{0.7mm}+\\
&+\int_{\Omega_0} \left|\hspace{-0.5mm}\left|\chi_\dta(Z\upsize) - \chi_\dta(Z)\right|\hspace{-0.5mm}\right| d\mu_{\Omega}\leq \epsilon_N \xrightarrow{N\rightarrow +\infty}0.\\
&\phantom{.}\qquad\qquad\qquad\qquad\qquad\qquad\qquad\qquad\qquad\qquad\phantom{.}\hspace{0.2mm}\blacksquare 
\end{align*}

\end{proof}

\subsection{Results on the Product $\class_\dta$ and the IMRP $Z_{\class_\dta}(t)$}
\label{app:onetoone}

Consider a single agent of class $\class = (\cstsp,\ctrsp)$ in a population model  $\pop\upsize = (\class, \ptrsp\upsize, \pinst\upsize)$, and a timed property $\dta = \dta(T) = (T, \lbset, \atprop,  \mathcal{C}\mathcal{C}, \gstsp,\ginst, \gfinst, \gtr)$. Let $\Sigma_{\class, \dta, T}$ be the set of time-bounded paths of $\class$ accepted by $\dta$ introduced in Sec. \ref{subsec:properties}, and let $Z\upsize(t)$ and $Z(t)$ be the two stochastic processes defined for the Fast Simulation in Sec. \ref{subsec:fastsim}. Moreover, consider the product $\class_\dta$ of agent class $\class$ and property $\dta$, and the IMRP $Z_{\class_\dta}(t)$, both introduced in Sec. \ref{sec:FMC}.

Let $AccPath(\class_\dta, T)$ be the set of time bounded paths of total duration $T$ that are accepted by $\class_\dta$ (i.e. such that they terminate in one of the final states of $\class_\dta$). Then, the result that guarantees that $\Sigma_{\class, \dta, T}$ is in 1:1 correspondence with $AccPath(\class_\dta, T)$ is just an adaptation to $\Sigma_{\class, \dta, T}$ and $AccPath(\class_\dta, T)$ of Lemma 3.9 in \cite{katoen}.

Finally, the fact that $P(T) = Prob_{Z_{\class_\dta}}\{AccPath(\class_\dta, T)\}$ comes from a (non-trivial) adaptation to time-dependent rates of Theorems 3.10 and 4.3 also in \cite{katoen}. We plan to provide the details of this formal proof in the future journal version of this paper.

\end{document}